# Extended characterization methods for covalent functionalization of graphene on copper

Petr Kovaříček,[1] Vladimír Vrkoslav,[2] Jan Plšek,[1] Zdeněk Bastl,[1] Michaela Fridrichová,[1] Karolina Drogowska,[1] Martin Kalbáč[1,*]

[1] J. Heyrovsky Institute of Physical Chemistry of the Academy of Sciences of the Czech Republic, Dolejškova 2155/3, 182 23 Praha, Czech Republic

[2] Institute of Organic Chemistry and Biochemistry of the Academy of Sciences of the Czech Republic, Flemingovo náměstí 542/2, 166 10 Praha, Czech Republic

| Author | ResearcherID |
| --- | --- |
| Petr Kovaříček | B-8431-2011 |
| Vladimír Vrkoslav | B-2146-2012 |
| Jan Plšek | F-5269-2014 |
| Zdeněk Bastl | H-6747-2014 |
| Michaela Fridrichová | A-7017-2013 |
| Karolina Drogowska | H-8445-2014 |
| Martin Kalbáč | F-5129-2014 |

## Abstract

Graphene is a material of great potential in a broad range of applications, for each of which specific tuning of the material's properties is required. This can be achieved, for example, by covalent functionalization. We have exploited two protocols for surface grafting, either by diazonium salts or by nucleophilic exchange, to perform graphene covalent modification directly on a copper substrate, which is routinely used for the synthesis of the material, and investigated the difference in reactivity compared with other substrates. The successful functionalization was confirmed by Raman and surface-enhanced Raman spectroscopy, mass spectrometry, X-ray photoelectron spectroscopy and scanning electron microscopy with energy-dispersive X-ray spectroscopy. In addition, we have found that the copper substrate can serve as a plasmonic surface enhancing the Raman spectra. Furthermore, the covalent grafting was shown to tolerate the transfer process, thus allowing *ex post* transfer from copper to other substrates. This protocol avoids wet processing and enables an all-gas-phase transformation of functionalized graphene, which eliminates the main sources of contamination.

[*] martin.kalbac@jh-inst.cas.cz, +420-266 05 34 08

# 1. Introduction

In the last decade, graphene has been identified as a unique material with remarkable properties.[1,2] Research groups around the globe have engaged themselves in the development of practical applications of graphene, one of which is sensor fabrication.[3–7] On the way to achieve this goal, two critical challenges arise: first, specific features for recognition and signalling have to be implemented in the material and second, accurate and unambiguous analysis of the material must be available to provide solid ground for drawing structure–functionality relationships.

Functionalization of graphene by covalent grafting represents a beneficial approach as it can simultaneously deliver signalling,[5] due to the graphene bandgap modulation and doping, as well as recognition via complementarity of intermolecular interactions.[8–10] High-quality large-area monolayer graphene is prepared by the chemical vapour deposition (CVD) method[11–13] on copper foil and then typically transferred onto another substrate (e.g. Si/SiO$_2$ wafers) by copper etching/polymer-assisted techniques[14] to perform the functionalization and characterization. However, this procedure is the main source of surface contamination caused by etchant, polymer and reactant residuals and solvent impurities.[14–16] The contaminants are typically local and they thus prevent homogeneous functionalization of the surface or, when in high amounts, contaminants can even completely prevent the reaction at graphene. Therefore, performing the functionalization reactions directly on copper would be highly desirable as it allows bypassing the etching and polymer-assisted transfer and performing the whole synthesis and functionalization sequence in the gas phase. In this way, wet processing is avoided and the fabrication can be implemented into a continuous line processing. Nevertheless, so far most of the reactions of CVD-grown graphene were performed on Si/SiO$_2$ substrates[17–32] as graphene on this support can be easily identified optically and/or by Raman spectroscopy. It has been shown that the substrate plays a significant role in covalent functionalization as it has an effect on the reactivity of the graphene monolayer,[28,33–35] therefore to achieve successful functionalization for a given substrate specific reaction conditions must be found.

The development of the functionalization protocols on substrates different from Si/SiO$_2$ is complicated because characterization methods for graphene or functionalized graphene are limited in number, sensitivity and experimental conditions. Therefore, there is an urgent need for extended characterization methods for on-surface chemical transformations. Among others, surface-enhanced Raman spectroscopy (SERS)[30,36–39] is particularly promising as it reveals the characteristic vibrational bands of the grafted moieties. In this case, performing the functionalization directly on copper can be again of benefit because copper has been reported as a SERS-active substrate[40] that allows the measurement of enhanced spectra without additional metal film deposition, which in turn makes this technique non-destructive.

We have performed functionalization of CVD graphene directly on the copper substrate used for the graphene synthesis and compared it with functionalization of graphene on a Si/SiO$_2$ wafer. To achieve graphene functionalization we exploited the protocols using either diazonium salts (Meerwein arylation)[24,29,31,41–43] or nucleophilic substitution on fluorinated graphene.[19,21,30,44,45] Raman spectroscopy is routinely used to investigate graphene functionalization due to the emergence of the D mode attributed to creation of $sp^3$ 'defects' in the 2-D $sp^2$ carbon lattice.[46–48] However, this feature is not indicative of the actual chemical nature of the functionalization product. In this vein, we have employed SERS,[30,36–39] mass spectrometry (MS),[33,49,50] X-ray photoelectron spectroscopy (XPS)[51] and scanning electron microscopy (SEM) with energy-dispersive X-ray spectroscopy (EDX)[52,53] to study graphene functionalization directly on the copper substrate.

## 2. Results

Single-layer graphene was synthesized using the CVD method on copper foil according to the standard procedure.[54] This material was either directly subjected to chemical modification, or, for comparison, transferred onto a silicon wafer (Si/SiO$_2$, 300 nm oxide layer) using the polymer-assisted method with nitrocellulose[14] and then reacted (Figure 1). For covalent chemical functionalization, we have selected two complementary approaches: direct Meerwein arylation with diazonium salts (method D)[29,31,41–43] or primary activation by fluorination followed by a nucleophilic exchange of fluorine atoms (method S).[21,30,45] In brief, diazonium salts were employed as approx. 5 mM solutions in deionized water (>18 MΩ cm$^{-1}$) and the solution was prepared either by dilution of commercial diazonia or by diazotization of aromatic amines by sodium nitrite in the presence of an acid (see SI for details). The fluorination was performed using XeF$_2$ in the gas phase as reported earlier.[30,55] The subsequent nucleophilic exchange by S-, N- and O-nucleophiles was performed either in the gas phase for volatile reagents or in solution. To show the general applicability of graphene functionalization on copper, we have performed a series of substitutions by altering the nucleophile for fluorine exchange using thioacetic acid, benzylthiol, octadecylthiol (S-nucleophiles), propylamine, 1-pyrenemethylamine (N-nucleophiles) and sodium ethoxide (O-nucleophile) or by employing various diazonium salts (4-nitrobenzenediazonium or 4-sulfonylbenzenediazonium). The corresponding experimental conditions are given in the SI.

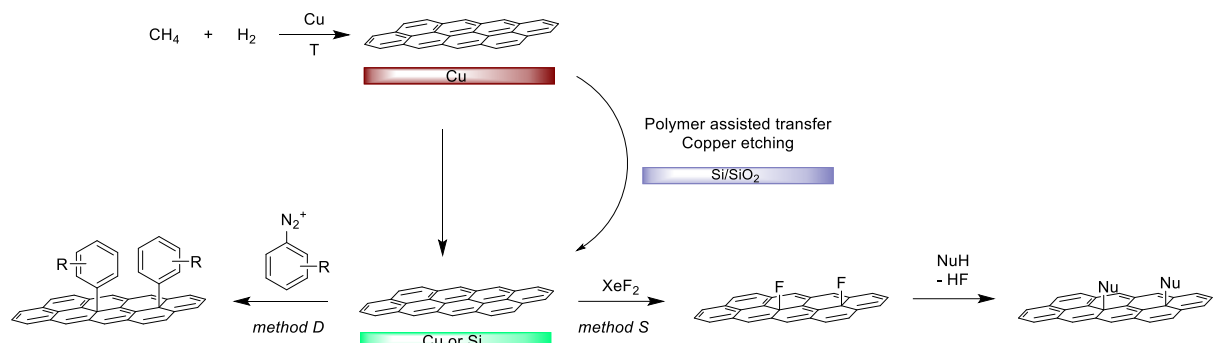

*Figure 1 Graphene was synthesized using the CVD method on copper foil, from which it can be transferred onto other substrates by polymer-assisted methods and copper etching. Graphene both on copper and Si/SiO$_2$ was covalently functionalized either by direct grafting of diazonium salts (method D) or activated by fluorination in the gas phase by XeF$_2$ followed by nucleophilic substitution (method S). Methods D and S are complementary in chemical compatibilities and in medium (solution or gas-phase processing).*

As a benchmark reaction, we have taken the previously studied system of single-layer CVD-grown graphene, which after fluorination was treated by thiophenol in the gas phase to undergo nucleophilic substitution of fluorine atoms by sulfur.[30] In this vein, we have performed the reaction sequence, *i.e.* fluorination and substitution by thiophenol (*g*), in parallel on both substrates and investigated them by Raman spectroscopy (Figure 2). Pristine graphene on Cu measured with 633 nm excitation laser shows the characteristic G and 2D bands at 1586 and 2658 cm$^{-1}$, respectively. Fluorination leads to the appearance of the D mode at 1343 cm$^{-1}$, which is accompanied by a dramatic decrease in the 2D mode intensity and shift of the G band to 1596 cm$^{-1}$. After the reaction with thiophenol in the gas phase, the 2D/D intensity ratio increases and the D′ band (1621 cm$^{-1}$) is resolved from the G mode, which shifts to lower wavenumber (1584 cm$^{-1}$). Interestingly, in the case of functionalization on copper, the characteristic phenylsulfanyl bands (476, 694, 1001, 1027 and 1083 cm$^{-1}$) have been observed in the spectrum due to the plasmonic enhancement by the substrate, while on Si/SiO$_2$ these bands appeared only after deposition of silver film and measuring using SERS (Figure 2).[30]

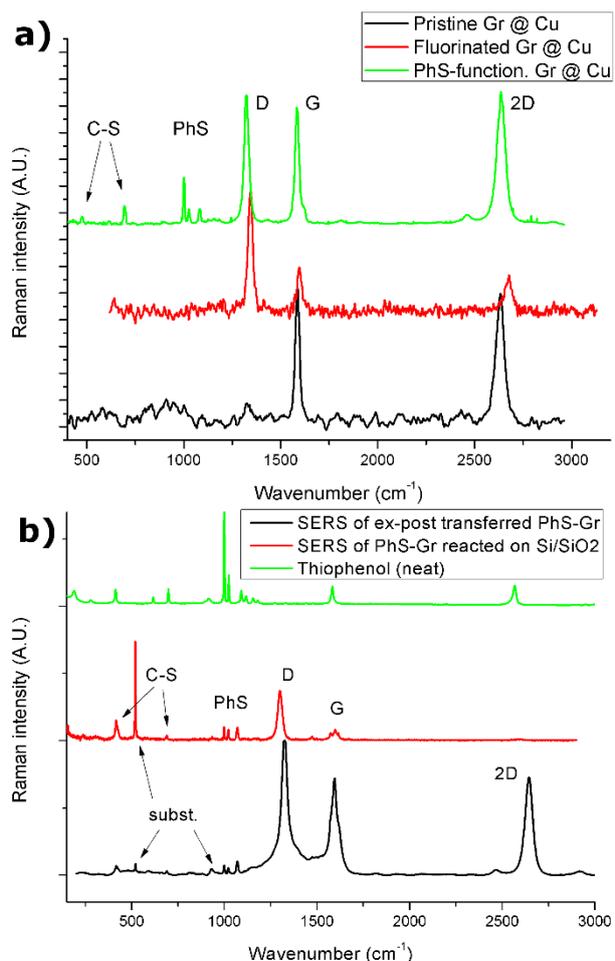

*Figure 2 Raman spectra of graphene covalent grafting by thiophenol on the copper substrate after each step. a) From bottom to top: black trace = as-synthesized graphene on Cu. Red trace = graphene on copper after fluorination. Green trace = functionalization with phenylsulfanyl groups leads to a decrease of the D mode and increase of the 2D mode intensities and resolving the D′ mode (1621 cm$^{-1}$). Importantly, combined in-plane C–H vibrations of the phenylsulfanyl group are visible as bands at 1001, 1027 and 1083 cm$^{-1}$, as well as C–S stretching (694 cm$^{-1}$) and deformation (476 cm$^{-1}$) bands. b) From bottom to top: black trace = graphene prepared and reacted on copper, then transferred onto Si/SiO$_2$, 12.5 nm silver film deposited on top and SERS spectrum measured showing the characteristic phenylsulfanyl vibrations as when measured directly on copper. Red trace = comparison with graphene functionalized on Si/SiO$_2$ substrate and measured with SERS showing identical characteristic vibrations as in the case of ex post transferred functionalized graphene. Green trace = comparison with neat thiophenol Raman spectrum.*

The reactive nucleophile can be also prepared *in situ*, such as, for example, in the case of octadecylthiol, prepared from the corresponding isothiouronium salt. The Raman spectrum measured on graphene@Cu showed the typical graphene modes (1327, 1584, 2643 cm$^{-1}$ for D, G and 2D, respectively) as well as characteristic vibrations of the octadecylsulfanyl moiety: 746 (C–S stretching), 524, 1062, 1102, 1130 (skeletal vibrations, CCCC trans and gauche), 1294 (twisting), 1440, 1461 (C–H deformation, scissoring), 2848 and 2884 cm$^{-1}$ (C–H stretching).[55] Importantly, we have intentionally damaged the sample by bending the copper foil and in the Raman spectra none of these characteristic signals was observed at the area without a graphene layer. A Raman map of a 15 μm × 15 μm area was processed and fitted so that the 2D mode and methylene bands (2848 and 2884 cm$^{-1}$) were extracted. Comparison of their intensities with the bright-field optical image showed that these signals are fully colocalized, which in turn shows that the octadecylsulfanyl functionalization proceeds solely on graphene. Moreover, all the functionalization protocols of graphene on copper were also performed simultaneously on the bare substrate and in none of the cases could any signal in the spectrum be identified (see SI for experimental traces).

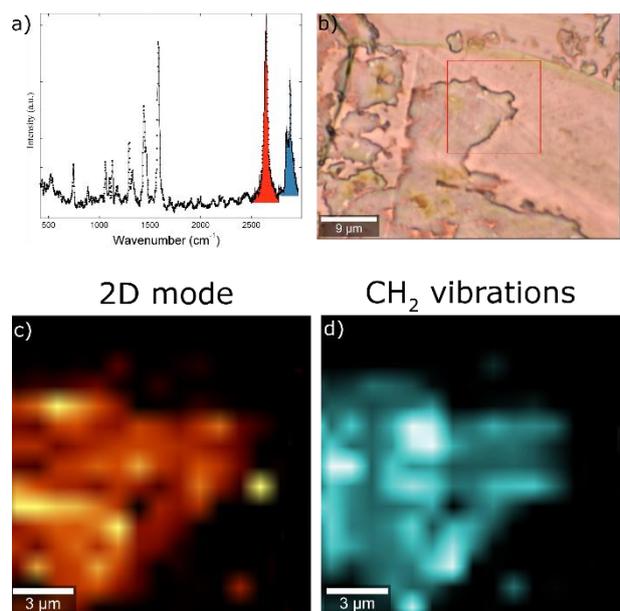

*Figure 3 Raman spectroscopy studies of the octadecylsulfanyl grafting on copper. a) Raman spectrum measured on graphene shows all important vibrations of the $C_{18}H_{37}S-$ moiety and graphene's characteristic D, G and 2D modes. b) Optical image of the sample where the graphene layer was damaged and discontinued, red square indicates the 15 μm × 15 μm area subjected to Raman mapping. c) Raman map of the 2D mode intensity over the selected area. d) Raman map of the CH$_2$ stretching vibrations intensity over the selected area. It can be clearly seen that the 2D mode and C–H vibrations are colocalized.*

We have used two MS methods, surface-enhanced laser desorption ionization/time-of-flight analyser (SELDI) and thermal programmed desorption–electron ionization (TPD–EI), and studied the functionalized graphene samples on both copper and Si/SiO$_2$ substrates. Graphene has been previously shown to serve as an efficient matrix for MALDI measurements[49,50] and we thus reasoned that this approach might be suitable also for characterization of graphene functionalization. With SELDI in negative mode, the functionalization by 4-sulfonylphenyl groups (method D), which did not provide an enhanced Raman signal on copper gave the molecular [M–H]$^-$ ion mass clearly detected at 157.0 *m/z*, which was also accompanied by graphene fragments' signals (equidistant signals at 84, 96 and 108 *m/z*, Figure 4a). The benchmark functionalization by phenylsulfanyl groups gave a signal at 109.0 corresponding to [M–H]⁻ of thiophenol (PhS⁻, Figure 4b) accompanied with signals of its oxidation products (up to PhSO$_3^-$, 157.0). The remarkable sensitivity tempted us to probe the imaging mode of the SELDI method and thus map the distribution of the grafted species over an extremely large area of about one millimetre size. As seen in Figure 4c, the signal of the PhS⁻ ion was indeed mapped at four selected rectangles with only a slight variation of its intensity within each of these areas, thus indicating rather homogeneous coverage of the surface by the grafted species.

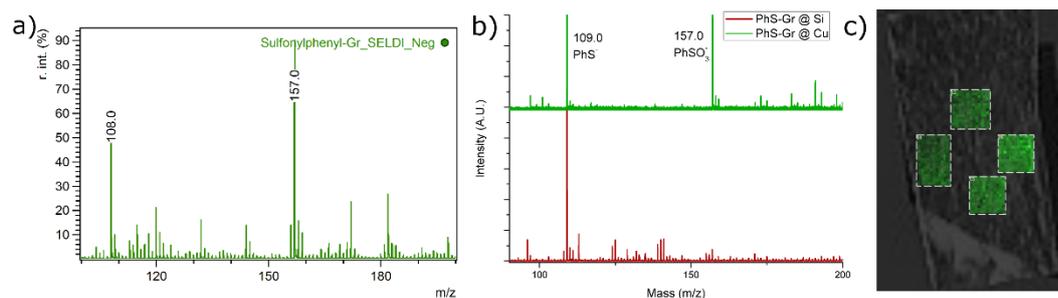

*Figure 4 SELDI MS of functionalized graphene. a) The spectrum of 4-sulfonylphenyl-functionalized graphene with the dominant [M–H]⁻ peak. b) SELDI spectra of phenylsulfanyl-grafted graphene recorded on copper (green) and Si/SiO$_2$ substrate (red) show the dominant signal belonging to the C$_6$H$_5$S⁻ (109.0) ion together with the product of its oxidation to sulfonate (157.0). c) SELDI imaging of the phenylsulfanyl-grafting distribution at selected areas subjected to mapping. The molecular ion is detected all over the surface with minor intensity variations.*

In a similar fashion, we have further exploited the SELDI method and investigated a series of various graphene graftings using both D and S functionalization approaches, which allowed us in all cases to identify the molecular ion of the covalently attached species always as negatively charged [M–H]⁻ ions. The results are summarized in Table 1 (see also SI for spectra and maps).

*Table 1 Summary of SELDI MS results from various covalent functionalizations with different reagents. Gr denotes graphene.*

| Method | Functionalization | Substrate | Detected mass |
|---|---|---|---|
| S | Phenylsulfanyl-Gr | Cu | 109.0 [M–H]⁻, 157.0 [M+3O–H]⁻ |
| S | Phenylsulfanyl-Gr | Si | 109.0 [M–H]⁻ |
| S | Benzylsulfanyl-Gr | Cu | 123.0 [M–H]⁻, 155.0 [M–H+2O]⁻, 171.1 [M–H+3O]⁻ |
| S | Octadecylsulfanyl-Gr | Cu | 285.3 [M–H]⁻ |
| S | Thioacetyl-(S)-Gr | Cu | 75.0 [M–H]⁻, 91.0 [M–H+O]⁻, 93.0 [M–H+H$_2$O]⁻ |
| S | 1-Pyrenemethylamino-Gr | Cu | 230.2 [M–H]⁻, 215.1 [M–NH$_2$]⁻ |
| D | 4-Sulfonylphenyl-Gr | Cu | 157.0 [M–H]⁻ |
| D | 4-Nitrophenyl-Gr | Cu | 122.0 [M–H]⁻ |

A very different situation, however, was observed when the phenylsulfanyl-grafted graphene was investigated using the TPD MS method. As the temperature increased, a strong peak at 78 *m/z* appeared corresponding to benzene as a thermal degradation product of phenylsulfanyl functionalization at both Cu and Si/SiO$_2$ substrates. The signal started to emerge at around 350 °C and reached the maximum amplitude at 425 °C. Importantly, no signal corresponding to the molecular ion could be detected up to 520 °C, thus clearly indicating a covalent bond of the phenylsulfanyl moiety to graphene, which can be further supported by the comparison of EI fragmentation patterns with published spectra databases (see SI).[†]

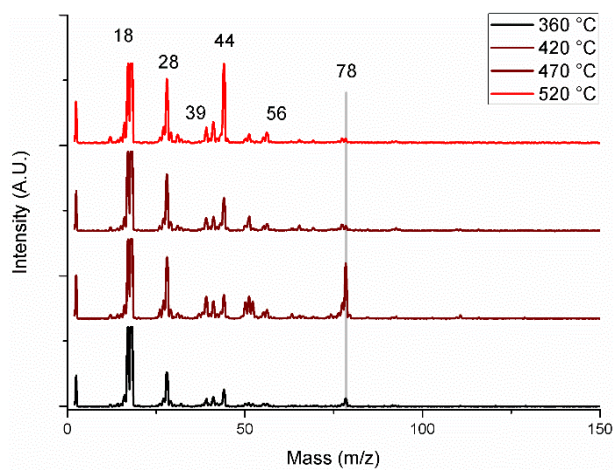

*Figure 5 TPD mass spectra of phenylsulfanyl-functionalized graphene on copper substrate recorded at indicated temperatures. The grafted species undergoes thermal decomposition to form benzene as the degradation product, which is then ionized and guided to the analyser. The fact that the molecular ion of thiophenol is not observed and that the fragmentation pattern is different indicates that the phenylsulfanyl moiety is indeed covalently bound to the monolayer graphene.*

TPD was measured for the substituents introduced by method S and in all cases we have identified the crucial molecules. The propylamino-functionalization gives a signal at 30 *m/z* corresponding to ethane (the second strong signal of ethane at 28 *m/z* overlaps with residual gases and thus is not suitable for investigation). Hydrogen sulfide (34 *m/z*) was detected due to the decomposition of thioacetyl-graphene. In the case of benzylsulfanyl-grafted graphene, a very intense peak at 91–92 *m/z* was observed corresponding to toluene and its fragment tropylium ion, which are overlapping in the spectra. The fact that equal 91 and 92 *m/z* signals appear in the spectrum agrees with the anticipated scenario that thermal elimination of the grafted species takes place.

---

[†] NIST Chemistry WebBook - Benzenethiol.
http://webbook.nist.gov/cgi/cbook.cgi?Name=benzenethiol&Units=SI&cMS=on#Mass-Spec (accessed February 2, 2016)

Molecular ions were detected for the thioacetate grafting which has also provided two hydrogen sulfide peaks at two distinct temperatures (200 vs. around 520 °C), which suggests that the thioacetate was partly hydrolysed prior to the thermal degradation. From the TPD profiles, it can be seen that the thermal decomposition occurs at different temperatures for each functionalization. While the signal of benzene, arising from the thermal decomposition of the phenylsulfanyl-grafted graphene, reaches its maximum at around 425 °C, the signal of toluene–tropylium ion starts to emerge at 250 °C and the signal of ethane (propylamine grafting) even as low as 150 °C.

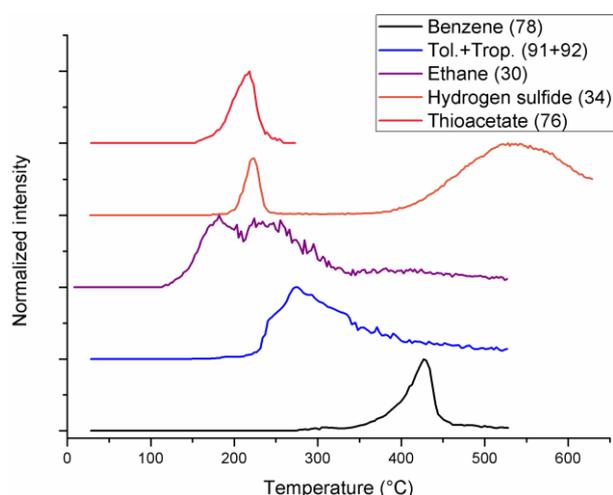

*Figure 6 TPD analysis of graphene covalent grafting. Temperature profiles for different functionalizations indicate the decomposition temperature at which significant fragments of the grafted moieties are detected: benzene (78) as a degradation product of phenylsulfanyl grafting, toluene–tropylium ion (91 + 92) coming from benzylsulfanyl, ethane (30) from propylamine and hydrogen sulfide (34) from thioacetic acid (76) grafting.*

From the XPS data, the graphene fluorination on copper under the conditions used gives about 15% fluorine atoms vs. carbon when measured shortly after the reaction and decreases in time based on the conditions, as has been described previously.[34,56] When the freshly fluorinated graphene on copper was immediately reacted with thiophenol, the surface composition determined using XPS was found to be $C_{1.00}F_{0.04}S_{0.15}$ (Figure 7a), suggesting that about 80% of fluorine was replaced by sulfur, which is in agreement with previous measurements on $Si/SiO_2$.[30]

To eliminate the influence of the substrate on the XPS measurement, we have first evaporated 12.5 nm thick Ag film onto the functionalized graphene sample on copper and then spin-coated a self-standing poly(methyl methacrylate) (PMMA) film on it. The silver film plays a dual role in this setup as it a) enables direct SERS measurement of the sample (see SI) and b) it effectively attenuates the photoelectrons emitted from the PMMA support, which would otherwise overwhelm the graphene carbon signal. Finally, the thick PMMA film also allows for peeling off the foil without copper etching by FeCl3, which caused problems with this functionalization, probably due to the oxidative properties of $Fe^{3+}$ (see SI for details). Indeed, the measured C 1*s* signal on the Ag:PMMA support is very similar to that measured on Cu showing only a small contribution of PMMA to its intensity (Figure 7b, fitted signals at around 286 and 288 eV corresponding to C-S, C–O and O-C=O species,[57–59] respectively). The small contribution of the support thus led to surface C:S ratio of $C_{1.00}S_{0.07}$ with S 2*p* spectra close to that on Cu substrate. It is noteworthy that in both cases part of the sulfur is oxidized to sulfone or sulfonate, as evidenced by the minor component at around 168 eV corresponding to 5 and 10% for Cu and PMMA support, respectively.

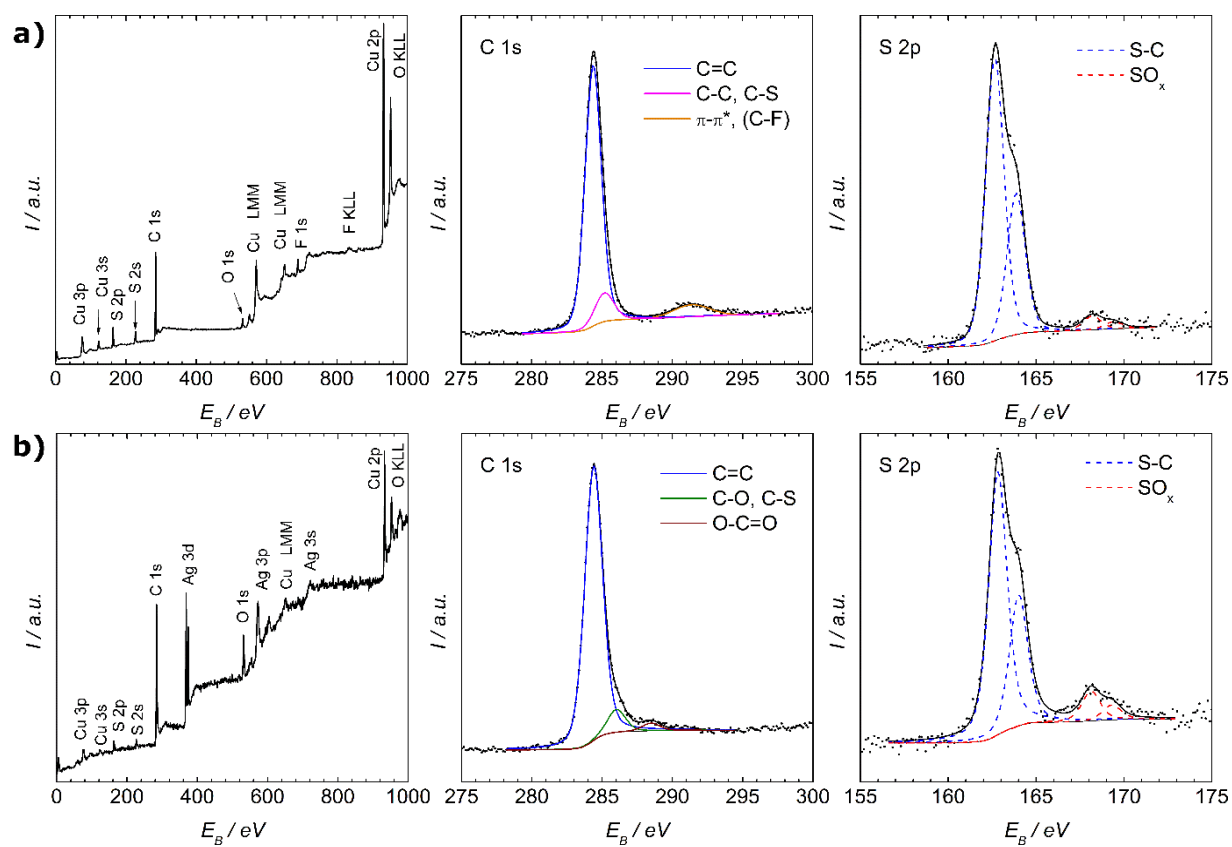

*Figure 7 XPS spectra of phenylsulfanyl-grafted graphene on a) copper and b) Ag:PMMA support. The silver layer efficiently attenuates the photoelectrons emitted from the PMMA support, thus giving a C 1s spectrum similar to that measured on copper. The S 2p$_{3/2}$ line located at around 163 eV is assignable to phenylsulfanyl species and the minor signal at around 168 eV can be attributed to its oxidation product (sulfone or sulfonate) ‡,[60–62] the line splitting is due to the spin–orbit coupling.*

The stoichiometry of the surface layer calculated from XPS spectra for other functionalization protocols are summarized in the SI Table 1. As reported earlier, the substrate has an effect on the reactivity of graphene,[28,33–35] giving about 5-10 % graphene

---

‡ NIST X-ray photoelectron spectroscopy database, NIST standard reference database 20, Ver. 4.1. Available at http://srdata.nist.gov/xps/ (accessed January 29, 2017)

carbon atoms functionalized on Si/SiO$_2$ and 10-15 % on copper substrate. The surface morphology and elemental distribution was further studied by SEM/EDX, which showed rather homogeneous coverage by the grafted species (except for thioacetic acid) and the corresponding images and spectra are provided in the SI.

## 3. Discussion

Complementarity of the two approaches used for graphene functionalization relies on processing and chemical compatibility. While Meerwein arylations proceed in solution and tolerate oxidative conditions, nucleophilic substitution can proceed both in solution and in the gas phase and tolerates reductive conditions. In the design of a particular graphene functionalization procedure, one thus has to choose the approach correctly to prevent undesired side reactions of the introduced moieties. The fact that the chemical grafting can be performed directly on the copper foil used for graphene CVD growth further broadens the compatibility range.

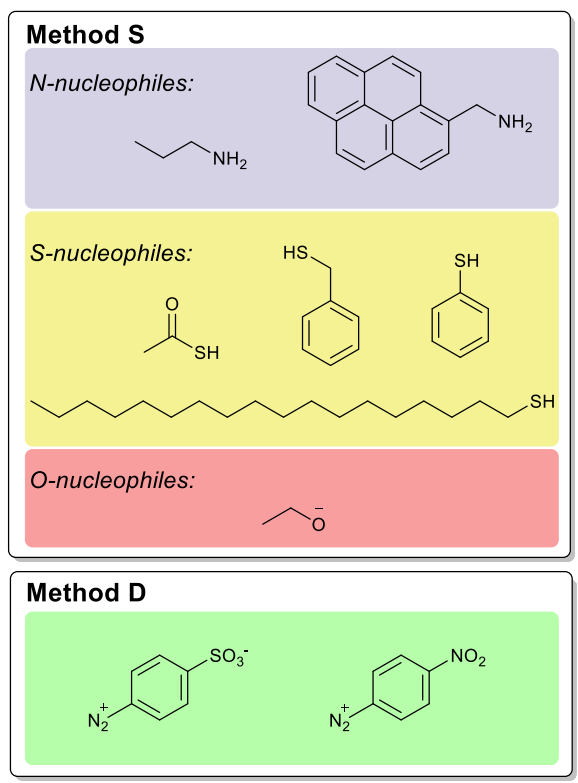

*Figure 8 Scheme of all the functionalization reagents used in the study. The substitution method was exploited with S-, N- and O-nucleophiles, the diazonium method used p-sulfonyl and p-nitrobenzene derivatives.*

The Raman signal surface enhancement in the presented cases is caused by the plasmonic metal film, which is either deposited on top of the sample (Ag), or, when functionalization is performed directly on copper, by the underlying substrate. In fact, SERS has been demonstrated for many metals, including copper, but its applications are hampered by its low stability towards oxidation under laser irradiation.[40,63] However, when graphene is present at the interface, it protects the metal from direct contact with the atmosphere and the imminent degradation is suppressed. This feature has nothing to do with the reported effect of graphene on long-term (anti)-corrosion of copper reported earlier.[64] It is noteworthy that partial surface corrosion is crucial for the observation of characteristic vibrations of graphene grafting on copper, such as in the case of fluorination ($XeF_2$) and consequent reaction with gaseous thiols. This procedure causes surface roughening, creating SERS-active hotspots and thus characteristic vibrations are observable even without additional deposition of metallic films. This is very important because evaporation of silver or gold layers on the sample makes it non-usable for further transformation or characterization. In this vein, performing graphene functionalization and acquiring SERS directly on the copper foil used to grow the material brings the benefits to react and to characterize non-destructively and without the polymer-assisted transfer step. Furthermore, one is not limited to the copper substrate, as *ex post* transfer is possible, and as shown by the phenylsulfanyl-grafted example also without the copper etching. On the other hand, the diazonium does not cause surface roughening and thus no characteristic Raman bands are observed. To visualize them, a silver film must be deposited on top and Ag-enabled SERS is recorded. In such a setup, characteristic vibrations are enhanced and can be attributed to products of both the Meerwein arylation and the azo coupling (see SI for details).

The Raman spectra and SERS can thus provide very important information about functional groups being grafted to graphene. However, it cannot distinguish whether or not the reactant maintained structural integrity, *i.e.* whether all atoms of the reactant were transferred to graphene or *vice versa*, if the material overreacted forming e.g. polymers commonly observed in reactions with diazonia.[65] By combination of the two MS methods used, several conclusions about the functionalization outcome can be drawn. First, detection of [M–H]⁻ ions using SELDI but not with TPD supports the conclusion of covalent grafting over noncovalent adsorption. Second, overreaction often encountered in diazonium grafting with relatively electron-rich reagents[65] can be unambiguously identified and discerned from monolayer functionalization. Third, mass spectra of thermal degradation products at given temperatures directly report on the chemical transformation taking place at the grafted surface. Finally, SELDI imaging can monitor the presence of the species of interest over very large areas, which are typically not accessible by other methods.

Among the analytical methods for surface chemistry, XPS is one of the very few that can provide not only a qualitative but also a quantitative description of the sample. In the benchmark reaction sequence, two heteroatoms are introduced to the graphene lattice: fluorine in the activation by $XeF_2$ and sulfur during the nucleophilic exchange by thiophenol. These elements give XPS spectra well separated from spectra of graphene carbon and substrate atoms, which is more suitable for precise quantification than rather difficult analysis of the complex C 1*s* peak shape. From XPS analysis, it has been found that approximately 80% of fluorine can be replaced by the nucleophile used, yet both the substrate and the nucleophile influence the reactivity. For example, in the thiol series, the amount of fluorine replaced correlated with the acidity of the sulfur nucleophile used, decreasing in the order thioacetic acid > thiophenol > benzylthiol. It was found that the reaction proceeds more on copper on which benzylthiol did react, unlike on the $Si/SiO_2$ substrate. Similarly, thioacetate substitution, which proceeded smoothly on $Si/SiO_2$, led to heavy corrosion of copper foil and rupture of the graphene monolayer. The choice of substrate can thus be regarded as another parameter that can be controlled to adjust the kinetics of the chemical reaction. Similar to sulfur or fluorine, nitrogen is also indicative in XPS, proving for example double reactivity of diazonia. Finally, if the substrate can bias the XPS analysis, *ex post* transfer can be performed using a thin metallic film, which attenuates hampering photoelectrons from the supporting material.

## 4. Conclusions

In this study, we have exploited two different protocols for graphene covalent functionalization: Meerwein arylation by diazonium salts and nucleophilic substitution of fluorinated graphene. In the latter, we have employed various S-, N- and O-nucleophiles to boost the pool of suitable reagents. The diazonium and substitution protocols are also largely complementary. We have demonstrated that the chemical modifications can be performed directly on the copper substrate used for graphene growth instead of the usual polymer-assisted transfer onto a silicon wafer. This approach not only eliminates the contamination by residual polymers and etchant, but it also allows measurement of surface-enhanced Raman spectra without the need for deposition of additional metal films on top, which makes the measurement of SERS easy, non-destructive and it allows direct confirmation of the successful functionalization of graphene. Moreover, the functionalized graphene can be transferred *ex post* on another substrate if desired. Finally, the graphene functionalization can be performed exquisitely in the gas phase, eliminating wet processing, which is beneficial for processability of the covalently modified material. We have supported our findings by several analytical methods, including Raman spectroscopy and SERS, MS, SEM/EDX and XPS.

Altogether, we believe that the presented work will promote the graphene chemistry field towards its development into applications.

## Acknowledgements

Authors would like to thank Ilona Spirovová for her help in processing XPS spectra. The work was supported by Czech Science foundation contract No.P208121062 and ERC-CZ project No. 1301. P.K. thanks the ASCR PPPLZ program for funding (L200401551). The authors acknowledge the assistance provided by the Research Infrastructure NanoEnviCz, supported by the Ministry of Education, Youth and Sports of the Czech Republic under Project No. LM2015073.

## References

**Supporting Information**. Experimental procedures, Raman spectra, SELDI spectra, TPD spectra and profiles, SEM/EDX images and spectra, blank experiments.